\documentclass[12pt,letterpaper]{article}
\usepackage[affil-it]{authblk}
\usepackage[left=.8in,right=.8in,top=.8in,bottom=.8in]{geometry}    
\usepackage{graphicx}
\usepackage{epsfig}
\usepackage{latexsym}
\usepackage{amssymb}
\usepackage{bm}
\usepackage{float}
\usepackage{subfigure}
\usepackage{amsmath}
\usepackage{amsfonts}
\usepackage{bm}
\usepackage{xcolor}
\newcommand{\be}{\begin{equation}}
\newcommand{\ee}{\end{equation}}
\newcommand{\bea}{\begin{eqnarray}}
\newcommand{\eea}{\end{eqnarray}}

\newcommand{\nn}{\nonumber}

\makeatletter
\g@addto@macro\@floatboxreset\centering
\makeatother

\usepackage[normalem]{ulem}

\begin{document}

%\title{Torsion and the probability of inflation}
%
%\author{Emma Albertini$^1$}
%\author{Stephon Alexander$^2$}
%\author{Gabriel Herczeg$^2$}
%
%
%
%\author{Jo\~{a}o Magueijo$^1$}
%\email{j.magueijo@imperial.ac.uk}
%\address{$^1$Theoretical Physics Group, The Blackett Laboratory, Imperial College, Prince Consort Rd., London, SW7 2BZ, United Kingdom}
%\address{$^2$Department of Physics, Brown University, Providence, RI, 02906, USA}
%
%\date{\today}

%%%%%%%%%%%%%%%%%%%%%%%%%%%%%

\title{Torsion and the probability of inflation}

\author[1]{Emma Albertini}
\author[2]{Stephon Alexander}
\author[2]{Gabriel Herczeg}
\author[1]{Jo\~{a}o Magueijo}
\affil{Theoretical Physics Group, The Blackett Laboratory, Imperial College, Prince Consort Rd., London, SW7 2BZ, United Kingdom}
\affil{Brown Theoretical Physics Center, Department of Physics,  Brown University, Providence, RI, 02906, USA}
\date{\today}

\maketitle

\begin{abstract}
We revisit the problem of the ``probability of inflation'' from the point of view of the Einstein-Cartan theory, where torsion can be present off-shell even in the absence of spinorial currents.  An informal estimate suggests that the barrier for tunneling from ``nothing'' into a classical universe becomes thinner and lower, should torsion be present even if only off-shell. This is confirmed by a detailed calculation, where the usual assumptions are re-evaluated and repurposed in our situation. Interestingly some approximations used in the literature (such as the WKB approximation) are not needed in general, and in particular in our case. When we consider wave packets centered around zero torsion, however, the conclusion depends crucially how these are built. 
With a Klein-Gordon current prescription for the measure and probability, for small torsion variance, $\sigma_c$, we recover the plane wave results. Nonetheless, for large $\sigma_c$, higher order corrections could reverse this conclusion.  
\end{abstract}

\maketitle

\section{Introduction}

The affine structure of General Relativity has been a source of interest right from its proposal. Even 
though the metric formulation gained more traction early on, the connection-driven Palatini and Einstein-Cartan formalisms were never completely
forgotten. With the advent of gauge theories, the situation almost reversed in some quarters, the connection gaining 
primacy over the metric when trying to import into quantum gravity the
non-perturbative quantization methods of non-abelian gauge theories. Most notably, in loop quantum 
gravity the Cartan connection morphs into an $SU(2)$ connection (the Ashtekar connection) via a canonical transformation. 

Naturally, once one accepts the independence of the connection with respect to the metric, the prospect of torsion and non-metricity 
is unavoidable. Thus, a ``landscape'' of possible theories emerges, including the Einstein-Cartan, teleparallel, and Palatini formulations. 
The implications for inflationary scenarios have been studied (see, for example \cite{Cartan-inf,Cartan-holst-inf}, 
\cite{Pal-inf,Pal-Higgs1,Pal-infl1,Pal-Higgs2} and  \cite{Telep-infl-withoutinfl,Telep-inf} for three distinct contexts). This applies to 
Higgs inflation, too~\cite{higgsinfl}. Curiously, as soon as a special and essentially unique inflationary model was selected by particle physics (the Higgs model), 
a plethora of gravitational alternatives sprawled out, in a sense undermining its predictivity~\cite{macarena}. 

In all this work, 
the effects of torsion were investigated within the setting of semi-classical inflation, but never 
within quantum cosmology. The last matter was a popular issue  in the 1980s: assessing the probability of inflation as inferred from quantum cosmology
with suitable boundary conditions~\cite{HH,Vil}. Different wave functions matching different boundary conditions once vied for a role as promoters of inflation
out of the quantum epoch. 

One therefore is led to wonder what could be the role of torsion within the quantum cosmology of inflation. In this we are helped by recent work~\cite{CSHHV}, where a close relation was found between  the wave functions obtained in the context of Ashtekar 
quantum gravity~\cite{kodama,realCS} and those in the metric formulation, such as the Hartle-Hawking~\cite{HH} and the Vilenkin~\cite{Vil} wave functions.  The latter are nothing but the Fourier transform of the so-called real Chern-Simons state, with suitable contours. 

Roughly speaking, metric driven formalisms are ``second-order,'' whereas connection based approaches are ``first-order,'' the terms referring to the order of their equations of motion (reflected in the number and structure of the constraints within the Hamiltonian formalism). First-order formulations are well-known to contain a parity violating sector associated with torsion~\cite{zlos1}, which can be particularly relevant in quasi-topological theories~\cite{Lee1,Lee2}. The parity-odd torsion appears in the Hamiltonian constraint, unlike the parity-even component (which is folded in the connection variable and only appears when trying to relate connection and metric via a Hamilton equation). Hence the Wheeler-DeWitt equation is sensitive to parity-odd torsion.

As it happens, it is straightforward to include the parity violating part of the torsion~\cite{zlos1} in the various wave functions previously considered~\cite{HH,Vil}, and even make wave packets with regularized norms by superposing such solutions~\cite{generalHawk}. The classical condition that the torsion is zero does not preclude an off-shell variance in the torsion. The presence of spinorial currents may add to the discussion.

Even without detailed calculations, it is easy to see that torsion should have an effect on the probability of inflation, particularly in the tunnelling approach. We present a back of the envelope calculation, which  reproduces the major features of our detailed calculation. 

\begin{figure}
\includegraphics[scale=.4]{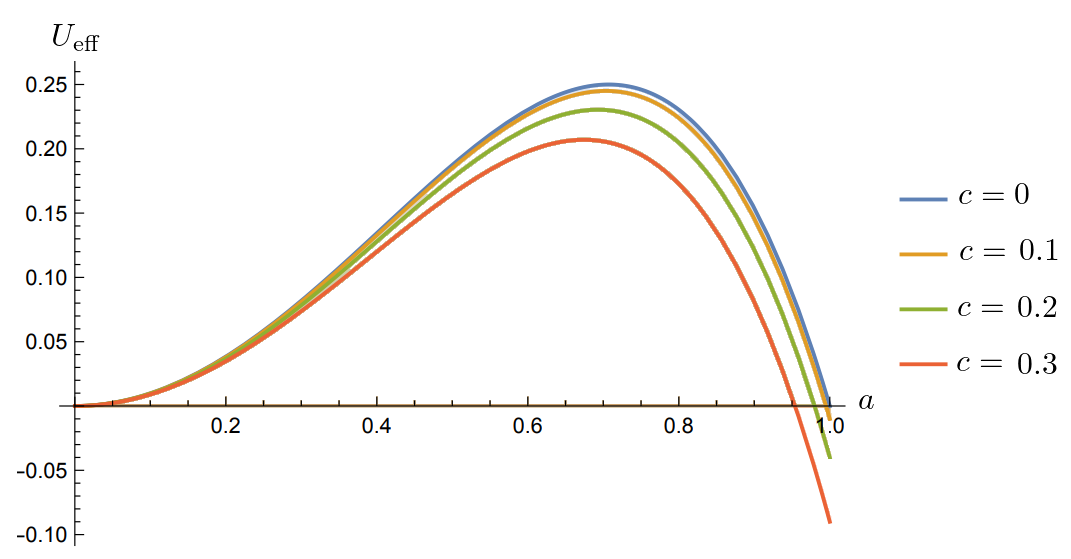}
\caption{Schematic depiction of the effective potential changing the torsion $c$.  }
\label{Uefffig}
\end{figure}

Following the approach of Vilenkin~\cite{Vil}, one can recast the Wheeler-DeWitt equation for Einstein gravity minimally coupled to a scalar field with potential $V(\phi)$ in the form of a Klein-Gordon-like equation
\be
\left(\nabla^2 - U_{\mbox{\scriptsize eff}}\right)\psi = 0. \label{WDW-KG}
\ee
Here
\be
\nabla^2 = G_{ijkl}\frac{\delta}{\delta h_{ij}} \frac{\delta}{\delta h_{kl}} + \gamma_{ij}\frac{\delta}{\delta h_{ij}} + \frac{1}{2}l_P^{-2}h^{-1/2}\frac{\delta^2}{\delta\phi^2} 
\label{laplacian}
\ee
is the Laplacian on superspace, with 
\be
G_{ijkl} = \tfrac{1}{2}h^{-1/2}\left(h_{ik}h_{jl} + h_{il}h_{jk} - h_{ij}h_{kl}\right)
\ee
being the DeWitt supermetric, $h_{ij}$ is the spatial metric on a Cauchy surface, and the coefficients $\gamma_{ij}$ depend on the choice of factor ordering. The effective potential is 
\be
U_{\mbox{\scriptsize eff}} = h^{1/2}\left[ -\,^{(3)}\!R + \frac{1}{l_P^2}\left( \frac{1}{2}h^{ij}\phi_{,i}\phi_{,j} + V(\phi)\right)\right]
\ee
where $^{(3)}\!R$ is the scalar curvature of the spatial metric. Reducing the Wheeler-DeWitt equation to minisuperspace amounts to restricting to spatial metrics of the form $h_{ij} = a(t)^2\delta_{ij}$, and scalar fields that are similarly homogeneous and isotropic, i.e. $\phi = \phi(t)$. If we also assume that the potential can be approximated by a cosmological constant $V(\phi) = \Lambda/3$, and look for solutions that don't depend on $\phi$ then the Wheeler DeWitt equation reduces to 
\be
\left(\frac{\partial^2}{\partial a^2} + \frac{p}{a}\frac{\partial}{\partial a} - U_{\mbox{\scriptsize eff}}(a)\right)\psi(a) = 0
\ee
where
\be
U_{\mbox{\scriptsize eff}}(a) = 4\left(\frac{3V_c}{l_P^2}\right)^2 a^2\left(k - \frac{\Lambda}{3}a^2\right)
\ee
with $p$ being a factor ordering parameter and $V_c$ the spatial volume of the Universe. If we take the Einstein-Cartan theory as the starting point rather than the standard metric formulation of Einstein gravity, the minisuperspace version of the Wheeler DeWitt equation has exactly the same form, except that now the effective potential obtains an additional dependence on the torsion \cite{generalHawk}
\be\label{potential}
U_{\mbox{\scriptsize eff}}(a, c^2)=4\left(\frac{3V_c}{l_P^2}\right)^2 a^2\left(k-c^2-\frac{\Lambda}{3}a^2\right)
\ee
where $c$ is the parity-odd component of the Cartan connection in homogeneous and isotropic models~\cite{zlos1}. Thus, the barrier between the ``nothing'' ($a=0$) and the
smallest classically allowed universe has width:
\be
L=\sqrt {\frac{3(k-c^2)}{\Lambda}}
\ee
and height:
\be
H=\frac{27V_c^2}{\Lambda l_P^2} (k-c^2)^2
\ee
(reached when $a=L/\sqrt{2}$).
Face value, the torsion has the effect of lowering and thinning the barrier, as we
illustrate in Fig.~\ref{Uefffig}. 
Standard quantum mechanics teaches us that for a square potential the transmission probability is a function of the height and width of the barrier, with:
\be
P\sim  \exp( -\sqrt H L)\sim
\exp\left(-\frac{9 V_c}{l_P^2 \Lambda}(k-c^2)^{3/2}\right)
\ee 
when the width and height are sufficiently large.

But is this true in detail? Clearly we cannot just lift the results in the literature without thought and serious adaptation. Such is the purpose of 
the present paper, 
which is organised as follows.
In Section II, we study the implication of leaving torsion off-shell on solutions of Wheeler-DeWitt with tunneling boundary conditions. The probability current is found to increase with torsion, both in the exact form and WKB approximation.\\
The discussion is extended in Section III, imposing the torsion to vanish classically but only as an expectation value at the quantum mechanical level. As wavepackets in torsion space are considered, the probability of tunneling gets suppressed.
The final Section compares the results and discuss remarks of the work.

\section{Monochromatic waves}\label{monochrom}
In this Section we start by uncritically repeating the calculation in~\cite{Vil} adding
on torsion to the wave functions. We then retrace our steps to evaluate the assumptions taken for granted and refine the calculation where appropriate, in view of the new element in the problem.

\begin{figure}
\includegraphics[scale=.5]{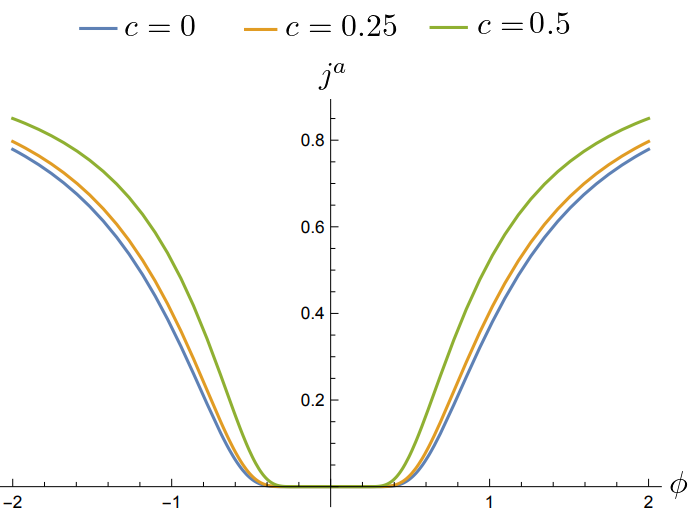}
\caption{The impact of torsion on the probability of large-field inflation, using the standard approximations. As intuitively predicted, torsion always increases the probability of inflation. (The scale of the $y$-axis in this plot is arbitrary.)}
\label{SmallField1}
\end{figure}

With the factor-ordering choice $p = -1$, the tunneling wave functions including torsion can be obtained in the same way as those for the Hartle-Hawking boundary conditions (see~\cite{zlos2,generalHawk} for details) and are
\begin{align} \label{Tunneling}
    \Psi_T= \mathcal{N}[ Ai(-z) +i \ Bi(-z)]
\end{align}
with
$$z= \left( \frac{9V_C}{\Lambda l^2_P} \right)^{\frac{2}{3}} \left(\frac{\Lambda a^2}{3}-k+c^2\right).$$
Throughout this calculation we may set:
\bea
k&\rightarrow&1\nn\\
V_C&\rightarrow&2\pi^2\nn\\
\frac{\Lambda}{3}&\rightarrow&\frac{V(\phi)}{8\pi G_N}
\eea
to mimic more closely the calculation in~\cite{Vil}, but we shall leave it general for future applications. As in~\cite{Vil}, we assume a scalar field dominated by its potential $V(\phi)$, so that the wave functions are those for deSitter space-time. However, the scalar field's kinetic term is essential for defining a conserved current and ultimately the probability of tunneling, even though it does not contribute to the wave function itself. In particular, associated with any given solution $\psi$ of the Wheeler-DeWitt equation \eqref{WDW-KG}, one can define a conserved current
\be
j^I = \frac{i}{2}(\psi^*\nabla^I \psi - \psi\nabla^I \psi^*) \label{generalCurrent}
\ee
where $\nabla_I$ is a covariant derivative on superspace with $\nabla^2 = \nabla_I\nabla^I$ as in \eqref{laplacian}, and indices are raised with an appropriate metric on superspace. In the minisuperspace picture (with $p = -1$), the current becomes simply
\bea
j^a &=& \frac{i}{2a}(\psi^*\partial_a\psi - \psi\partial_a\psi^*) \nonumber \\
j^\phi &=& -\frac{i}{2a^3}(\psi^*\partial_\phi\psi - \psi\partial_\phi\psi^*) = 0, \label{current} 
\eea
and $j^a$ can be interpreted as a probility density for the scalar field $\phi$.
As in the standard calculation~\cite{Vil},
we evaluate this in the classically allowed region close to the barrier (but ``not too close so that we can use the WKB approximation''). Using the WKB approximation in the classically allowed regions, we find
\bea
     \Psi_T&\approx& \frac{\mathcal{N} }{\sqrt{\pi}} z^{-\frac{1}{4}} \left[\text{sin}(\zeta+\frac{\pi}{4})+i \  \text{cos}(\zeta+\frac{\pi}{4})\right]\nn\\
     &=& \frac{\mathcal{N} }{\sqrt{\pi}z^{\frac{1}{4}}} i \ e^{-(\zeta +\frac{\pi}{4})} = \frac{\mathcal{N} e^{\frac{i \pi}{4}}}{\sqrt{\pi}z^{\frac{1}{4}}}e^{-i\zeta }
\eea
with $$\zeta= \frac{2}{3} z^\frac{3}{2}=\frac{6V_C}{\Lambda l^2_P}  \left(\frac{\Lambda a^2}{3}-k+c^2\right)^\frac{3}{2} .$$
It follows that
\begin{align}
    j^a\approx  -\mathcal{\Im} \frac{\Psi^*\partial_a\Psi}{a}  = |\Psi_T|^2 \frac{6 V_c}{l^2_P}\left(\frac{\Lambda a^2}{3}-k+c^2\right)^\frac{1}{2}.
\end{align}
But since $  |\Psi_T|^2= (\pi \sqrt{z})^{-1}|\mathcal{N}|^2$,
we have
\begin{align}
 j^a=  \frac{6 V_c}{l^2_P}\frac{|\mathcal{N}|^2}{\pi z^\frac{1}{2}}\left(\frac{\Lambda a^2}{3}-k+c^2\right)^\frac{1}{2}=\frac{2  \Lambda }{3\pi}\left( \frac{9 V_c}{l_P^2\Lambda }\right)^\frac{2}{3} |\mathcal{N}|^2. 
\end{align}
The crucial input here, therefore, is the normalization factor. This is fixed by the assumption $\Psi(a=0)=1$, and this is extensively justified in~\cite{Vil} (see Section IVB therein). 
Then we have
\begin{align}
    \mathcal{N}^{-1}=  Ai(-z_0) +i \ Bi(-z_0)
\end{align}
with 
$$z_0= \left( \frac{9V_C}{\Lambda l^2_P} \right)^{\frac{2}{3}} (c^2-k).$$
If we can assume $c^2\ll k=1$, we can use the WKB approximation again, this time in the classically forbidden region (i.e. $z_0 < 0$), to find
\begin{align}
    \mathcal{N}^{-1}= \frac{(-z_0)^\frac{-1}{4}}{\sqrt{\pi}} \left(\frac{e^{-\zeta_0}}{2}+i e^{\zeta_0}\right)
\end{align}
with 
$$\zeta_0= \frac{6V_C}{\Lambda l^2_P}  (k-c^2)^\frac{3}{2}.$$
Thus, the normalization needed for the current is of the form
\begin{equation}
   |\mathcal{N}|^2=\frac{\pi  (-z_0)^\frac{1}{2}}{ \frac{e^{-2\zeta_0}}{4}+e^{2\zeta_0} }  
\end{equation}
where the denominator is the sum of two exponentials, where in ``most cases'' (i.e., those where the probability ends up being small), the second one dominates.
Hence the final result is
\begin{align}\label{Prob1}
j^a \sim e^{-\frac{12 V_c}{l_P^2 \Lambda} (k-c^2)^\frac{3}{2}}
\end{align}
which is remarkably close to the one obtained via analogy and hand waving in the introduction. Therefore, our intuition seems vindicated by a rerun of the standard calculation, with its intrinsic assumptions. As Fig.~\ref{SmallField1} shows, torsion always increases the probability of inflation, as predicted in the Introduction.

We must, however, re-examine some of the assumptions made in the standard calculation to assess whether they break down here. 

\subsection{The effect of the approximations on ${\cal N}$}

In the above it was assumed that the tunneling probability is small. Since the effect of torsion is to increase this probability, eventually the approximation breaks down. An improvement on (\ref{Prob1}) in this respect is
\begin{equation}
    j^a \approx \frac{\sqrt{k-c^2}}
    {\frac{1}{4}e^{-\frac{12 V_c}{l_P^2 \Lambda} (k-c^2)^\frac{3}{2}}+e^{\frac{12 V_c}{l_P^2 \Lambda} (k-c^2)^\frac{3}{2}}}
    \label{currentWKB}
\end{equation}
where we have dropped numerical coefficients but {\it not} power-law factors of $\Lambda$ (they cancel), and have not assumed the domination by one exponential. We have used the WKB approximation but this should not be a problem because $z_0$ is not at a turning point.

\begin{figure}
\includegraphics[scale=.5]{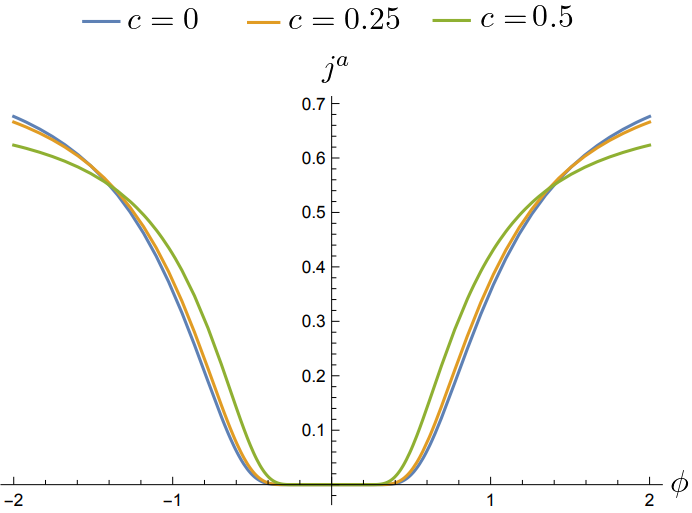}
\caption{The impact of torsion on the probability of large-field inflation, using a refinement (but still WKB) to the usual formula. As we see the refinement does affect the conclusions. }
\label{SmallField2}
\end{figure}

In Fig.~\ref{SmallField2} we recalculate the probability for a large field inflationary model, in this case a simple mass potential. This should be compared with the first row in Fig.4 of~\cite{Vil}). As we see, according to this refinement, it is not true that torsion always increases the probability of inflation. That only happens if the probability is not too large, i.e. in a large field inflationary model if we do not tunnel into a super-Planckian field, which is precisely what the model wants to do. As pointed out in~\cite{Vil}, the largest values of the probability correspond to super-Planckian values of the potential $V(\phi)>1$. The semi-classical approach taken might then not be trusted. 

But, then, in this regime, why do we use the WKB approximation?

\subsection{The effect of the WKB approximation}
\begin{figure}
\includegraphics[scale=.4]{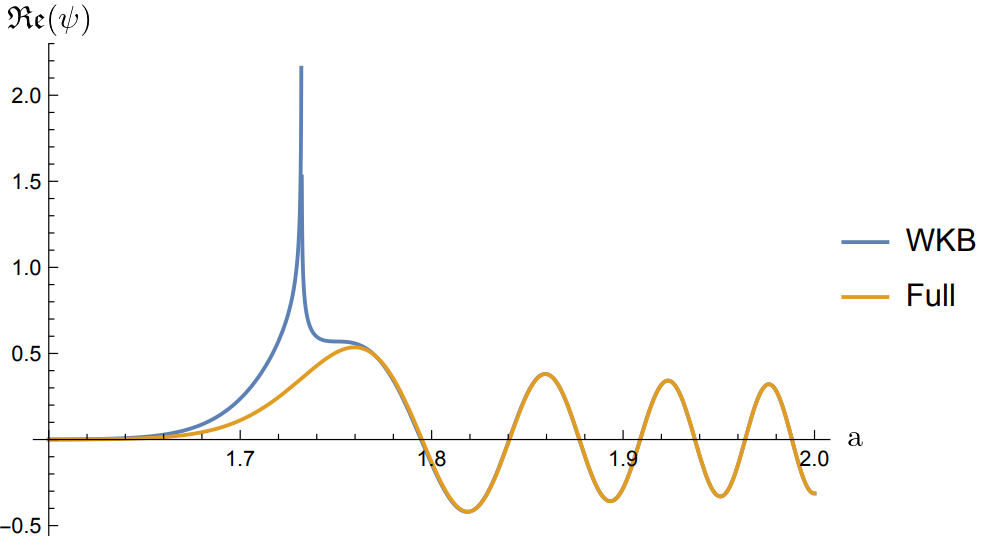}
\caption{The failure of WKB at the point into which the Universe tunnels. The currents associated with the two functions, however, are both constant and identical. The only difference results from the wave function normalization at $-z_0$ rather than at $z=0$.}
\label{WKBAi}
\end{figure}

Even before we get to the effects of torsion, 
one could have found it objectionable that 
the point into which the Universe tunnels out of nothing is a turning point between classically allowed and forbidden regions, where the WKB approximation is known to fail, and yet that is precisely the approximation used in the literature for evaluating the probability of tunneling. Indeed the WKB wave functions become divergent at the turning point, as illustrated in Fig.~\ref{WKBAi}, even though the associated current is finite. Why does this work? And what happens if we use the full wave functions in evaluating the current?

As it happens the fact that the current is a constant in $a$ is not an artifact of the WKB approximation, but follows from the Wronskian of the Airy function. Indeed, computing the current in terms of the Airy function using (\ref{current}), 
we obtain
\begin{align}
    j^a=-\frac{W(Ai(-z(a)),Bi(-z(a)))}{Ai(-z_0)^2 +\ Bi(-z_0)^2}
\end{align}
where the Wronskian is given by 
\be
W(Ai(-z(a)),Bi(-z(a)))=-\frac{\partial z}{\partial a}\frac{1}{\pi}.
\ee
Hence
\begin{align}
    \label{monocurrent}
    j^a=\frac{2 \Lambda }{3 \pi} \left( \frac{9V_C}{\Lambda l^2_P} \right)^{\frac{2}{3}} \frac{1}{Ai(-z_0)^2 +\ Bi(-z_0)^2}
\end{align}
which depends just on the normalization factors.
 Applying the WKB approximation to the denominator, the result in (\ref{currentWKB}) is precisely recovered. As the current is a constant as a function of $a$, the probability of tunneling can be calculated using WKB, as in the literature, although the approximation breaks down at the turning point $z=0$.
 
 The only refinement obtained by dropping the WKB approximation, therefore, is in the normalization of the wave function, i.e. in the use of exact results instead of WKB ones at $-z_0$. As we see 
 in Fig.~\ref{SmallField2} torsion now always increases the probability of inflation. Indeed, for large torsion, the barrier may become so thin, that $a=0$, where the wave function is normalized, is still in the regime where the WKB approximation breaks down. 
 
\begin{figure}
\includegraphics[scale=.5]{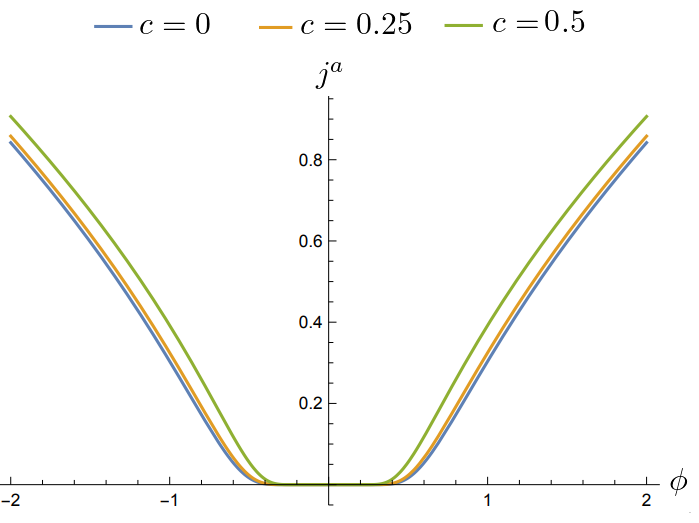}
\caption{The impact of torsion on the probability of large-field inflation, dropping the WKB approximation altogether. As we see torsion always increases the probability of inflation, even in the super-Planckian domain.  }
\label{SmallField3}
\end{figure}

\section{Torsion wave packets}\label{wavepacks}
So far we have considered the possibility of including non-vanishing torsion in our wave functions, with the result that torsion can increase the tunneling probability. This could be of use when torsion would be present classically, for example if it is sourced by spinorial currents. 
Otherwise, if torsion vanishes classically, one is faced with the problem of how to impose that condition quantum mechanically. This is a non-trivial problem because the torsion-free condition leads to second class constraints~\cite{zlos1}. Ordinarily, one would either solve the constraint and then quantize or quantize using Dirac brackets. Another option is to take an approach similar to the Gupta-Bleuler quantization and impose that the \emph{expectation value} of the torsion must be zero by building linear superpositions of solutions into wave packets centered around zero torsion~\cite{zlos2}.

Crucial to this approach is the measure and inner product used to make sure that the wave packets are normalized. As explained in~\cite{zlos2} the natural measure, emerging from the form of the plane wave solutions is $dc^2$. Indeed
the Airy functions are delta normalized in shifts in their arguments, which means $c^2$. Hence the natural measure for superpositions is
\be\label{measure1}
\psi_i(a) =\int dc^2 A_i(c^2) \psi(a,c)
\ee
with the inner product between two states defined as
\bea
\langle \psi_1|\psi_2\rangle&=&
\int dc^2 A^*_1(c^2)A_2(c^2) \label{innc}
\eea
(this generalizes beyond minisuperspace, as discussed in~\cite{realCS}). 

Solutions of the form \eqref{measure1} satisfy the Wheeler-DeWitt equation 
\be
\hat{H}\psi=
\left(\frac{\partial^2}{\partial a^2} + \frac{p}{a}\frac{\partial}{\partial a} - U_{\mbox{\scriptsize eff}}(a, c^2)\right)\psi = 0
\ee
where $U_{\mbox{\scriptsize eff}}(a,c^2)$ is given by \eqref{potential}, bearing in mind that $\hat H$ depends on the torsion $c^2$ to be seen as an operator. We use a representation diagonalizing $\hat c^2$ (as opposed to its conjugate momentum $p_c$), so $\hat c^2$ reduces to a real number applied to eigenstates of $\hat c^2$ but not for superpositions. Specifically,
\begin{equation}\label{c2noneig}
    \hat c^2 \psi_i(a) =\int dc^2 c^2 A_i(c^2) \psi(a,c).
\end{equation}
This has the implication of invalidating the argument leading to the conservation of the usual current \eqref{generalCurrent}. Indeed, from equations \eqref{WDW-KG} and \eqref{generalCurrent} we have
\be
\nabla_I j^I = \frac{i}{2}\left(\psi^* U_{\mbox{\scriptsize eff}}\, \psi - \psi\, U^*_{\mbox{\scriptsize eff}}\,\psi^*\right).
\ee
Thus, current conservation follows from the Wheeler-DeWitt equation \emph{and the fact that $U_{\mbox{\scriptsize eff}}$ is a real function}: it cannot have an imaginary part, which it does not in our case, or be an operator acting on a non-eigenstate, which it is. In view of (\ref{c2noneig}) we have
\bea
\partial_I j^I&\propto &\psi^*\!\!\int dc^2 c^2 A_i(c^2) \psi(a,c) - c.c.
\eea
In contrast, the alternative current
\be 
j^I(a) =\int dc^2 A(c^2)j^I(a,c^2)
\label{conservedcurrent}
\ee
is conserved in general for superpositions (\ref{measure1}) and reduces to the usual current, Eq.~\eqref{current} for eigenstates of $\hat c^2$. In this expression $j^I(a,c^2)$ is the current for each fixed $c$ component. This current will therefore form the basis for our probability arguments using torsion wave packets. 
\section{Probability of inflation for a Vilenkin beam} \label{probinfl}
The idea, therefore is to impose zero-torsion quantum mechanically by taking packets of the form (\ref{measure1}) with $A(c^2)$ a Gaussian distribution in $c^2$, centered around $c^2=0$ with standard deviation $\sigma_c$. Hence we may consider a ``Hartle-Hawking beam''~\cite{zlos2}:
\be
\Phi_H(a^2) = \int dc^2 A(c^2) \Psi_H(a^2, c) \label{beam}
\ee
with $\Psi_H(a^2, c) \propto \textrm{Ai}(-z)$.
This integral can be performed with the help of a standard formula for the Airy transform of a Gaussian distribution. The Airy transform of a given function $f(x)$ is defined as: 

\be
\Phi_{\alpha}(y) = \frac{1}{\alpha} \int dx \,f(x) \textrm{Ai}\!\left(\frac{y-x}{\alpha}\right) \label{AT1}
\ee
which for a Gaussian distribution $f(x) = \frac{1}{\sqrt{2\pi}}e^{-x^2}$, becomes \cite{zlos2, Soares}
\be
\Phi_{\mathcal{G}(\alpha)}(x)= \frac{1}{|\alpha|} \exp\left(\frac{x}{4\alpha^3}+\frac{1}{96 \alpha^6}\right) \textrm{Ai}\!\left(\frac{x}{\alpha}+\frac{1}{16 \alpha^4}\right). \label{AT2}
\ee
In the case of the tunneling wave function, we are interested in computing the analog of of \eqref{beam}, but with $ \Psi_H(a^2, c)$ replaced by $\Psi_T(a^2, c)$, which is given by \eqref{Tunneling}. In order to again make use of the Airy transform \eqref{AT1}, \eqref{AT2}, we refer to the Airy function identity 
\be
\textrm{Ai}(e^{2\pi i/3}z) + i\,\textrm{Bi}(e^{2\pi i/3}z) = 2 e^{\pi i/3}\textrm{Ai}(z)
\ee
so that
\begin{align}
    \Phi_T(\Tilde{a}^2)= \tilde{\mathcal{N}}\!\! \ \exp\left(\frac{\Tilde{\sigma}^2_c}{2}(\Tilde{k}-\Tilde{a}^2+\frac{1}{6}\Tilde{\sigma}_c^4)\right) \left(\textrm{Ai}(-\tilde{z}) +i \ \textrm{Bi}(-\tilde{z})\right) \label{V-beam-normalized}
\end{align}
where \begin{align}
     \Tilde{z}=\beta(- k+ \frac{\Lambda a^2}{3}- \beta^3{\sigma}^4_c), \qquad \beta= \left(\frac{9V_c}{\Lambda l^2_P }\right)^{2/3},
\end{align}
$$\Tilde{a}^2= \beta \frac{\Lambda a^2}{3}, \qquad \Tilde{k}= \beta k, \qquad \Tilde{\sigma}_c= \beta {\sigma}_c, $$ and by virture of the tunneling boundary condition $\Phi_T(a = 0) = 1$, we have $$\Tilde{\mathcal{N}}^{-1}= \exp\left(\frac{\Tilde{\sigma}^2_c}{2}(\Tilde{k}+\frac{1}{6}\Tilde{\sigma}_c^4)\right) \left(\textrm{Ai}(-\tilde{z}_0) +i \ \textrm{Bi}(-\tilde{z}_0)\right)$$ with $\tilde{z}_0=\beta(- k- \beta^3{\sigma}^4_c)$.
\subsection{The probability current }
Using the definition of the current in \eqref{conservedcurrent} for monochromatic waves, we have an expression for the beam current as
\be
j^a_{beam} = \frac{2 \Lambda }{3 \pi} \left( \frac{9V_C}{\Lambda l^2_P} \right)^{\frac{2}{3}}\int dc^2 A(c^2) \frac{1}{\text{Ai}(-z_0)^2 +\ \text{Bi}(-z_0)^2}
\ee
where $$A(c^2)= \sqrt{\frac{1}{2 \pi \sigma_c^2} } \  \exp \left({ -\frac{c^4}{2 \sigma_c^2}}\right)$$
Firstly, we focus on the expression for the monochromatic current \eqref{monocurrent} and we Taylor expand around $c^2=k$, 
which 
\be
j^a = \sum\limits_{n = 0}^{\infty}j_n^a(c^2-k)^n
\ee
with the first few terms in the series being
\bea
j^a_0 &=& \frac{3^{1/3}}{2 \pi } \beta \Lambda \Gamma\! \left(\tfrac{2}{3}\right)^2 \\
j^a_1 &=& \frac{3^{2/3} \beta^{2} \Lambda \Gamma\! \left(\frac{2}{3}\right)^3}{2 \pi  \Gamma \! \left(\frac{1}{3}\right)} \\
j_2^a &=& 0 \\
j_3^a &=& \frac{3^{1/3}}{2 \pi} \beta^{4/3} \Lambda \Gamma\! \left(\tfrac{2}{3}\right)^2 \left(\frac{1}{3}-\frac{3\Gamma\! \left(\tfrac{2}{3}\right)^3}{\Gamma\! \left(\tfrac{1}{3}\right)^3}\right) \\
j_4^a &=& -\frac{\beta^{5} \Lambda \Gamma\! \left(\frac{2}{3}\right)^3 \left(36 \Gamma\! \left(\frac{2}{3}\right)^3-5 \Gamma\! \left(\frac{1}{3}\right)^3\right)}{3^{1/3}8\pi  \Gamma\! \left(\frac{1}{3}\right)^4} \\
&& \hspace{2.3 cm} \vdots \nonumber
\eea
and all terms of the form $j_{2 + 3n}^a$ vanish. The choice to expand around $c^2 = k$ is merely a matter of convenience, since the coefficients of the expansion take a particularly simple form in this case. However, given that the Gaussian smearing $A(c^2)$ is peaked around $c^2 = 0$, it would be somewhat more natural to consider expanding about $c^2 = 0$. Indeed, for an arbitrary choice of $k$, all terms in the expansion about $c^2 = k$ will contribute zero-order contributions to total current. With this in mind, we focus now on computing the current for the case $k=0$, and we find a general expression for the current of the beam given by
 \be
     j^a_{beam} = \frac{4  \Lambda \beta }{3 \pi} \sqrt{\frac{1}{2 \pi \sigma_c^2} }  \sum_{n=0}^{\infty}\frac{f^{2n}(0)}{(2n)!} I_{2n}\left(1/\sigma_c^2\right) \label{beamCurrent}
 \ee
where 
\be
I_n(u)= \frac{(n-1)!!}{2^{{n}/{2}+1}+u^{{n}/{2}}} \sqrt{\frac{\pi}{u}}
\ee and 
\be
f^{n}(c^2)= \frac{\partial^n}{\partial {c^2}^n} \left(\frac{1}{\text{Ai}(-z_0 )^2+\text{Bi}(-z_0)^2 }\right). 
\ee
When the Gaussian smearing is sharply peaked---i.e., when $\sigma_c$ is small---the first few terms in the expansion will be a good approximation for the total current. At zero order, we get
\be
j^a_{0, beam} =  \frac{3^{1/3}}{2 \pi } \beta \Lambda \Gamma\! \left(\tfrac{2}{3}\right)^2 \label{j^a_0}
\ee 
 recovering the result for a monochromatic wave with $c=0$ and $k=0$. When $k=0$, only even terms in the expansion contribute to the total current for the beam, so the next non-vanishing contribution comes from the fourth-order term in the series \eqref{beamCurrent}: 
 \be
j^a_{4, beam} =\frac{3^{1/6}  \sqrt{\pi } \beta^2 \Lambda \sigma_c^4 \left(144 \pi^2 -5 \sqrt{\pi } \Gamma\! \left(\frac{1}{6}\right)^3\right)  }{2^{2/3}  \Gamma\! \left(\frac{1}{6}\right)^5}. \label{j^a_4}
 \ee
Note that the zero-order contribution to the total beam current is independent of $\sigma_c$, and this is indeed the exact result in the limit as $\sigma_c \to 0$ where the Gaussian tends to a delta function. Meanwhile, the correction \eqref{j^a_4} is negative whenever $\Lambda > 0$, suggesting that beams that are sharply peaked around $c^2 = 0$ have a greater tunneling probability than those that are more uniformly distributed in $c^2$.

\section{Conclusions}
In this paper we  formalized the basic intuition that torsion increases the probability for the creation of a Universe out of nothing because it thins and lowers the barrier. We found that, when all approximations are suitably re-evaluated, this intuition is correct for monochromatic waves in the torsion, generalizing the usual Vilenkin wave function. It is interesting to note that we did not need to rely on the WKB approximation (as in past literature): in fact the Wronskian method provides a more robust derivation of many results found here (and in previous literature).

This result by itself is applicable in situations where torsion would be present classically, for example if it is sourced by a spinorial current. It might also be true when the classical condition $c=0$ is enforced quantum mechanically with $\langle c\rangle =0$ but $\langle c^2 \rangle >0 $. For example we could build packets using the measure
\be
\psi(a) =\int dc  A (c) \psi(a,c)
\ee
and then select an amplitude Gaussian in $c$ (and not in $c^2$ as in
(\ref{measure1})).
Then the results in Section~\ref{monochrom} are correct with 
\be\label{c2rep}
c^2\rightarrow \sigma^2(c),
\ee
at least qualitatively.

Unfortunately the one case where a standard unitary inner product has been found~\cite{zlos2} leads to a wave packet Gaussian in $c^2$, and not in $c$. This produces the so-called Hartle-Hawking beam~\cite{zlos2}, which we generalized to the Vilenkin wave function in Section~\ref{wavepacks}. However, the inner product argument fails for the Vilenkin beam. This is best seen in the connection representation, where instead of delta normalization we would get
\be
\langle  c'^2|c^2\rangle \propto \int db \, e^{ib (c^2-c'^2)} 
\ee
where the integral goes over the positive real domain and the negative imaginary axis~\cite{CSHHV}, instead of the real line, as in the case for Hartle-Hawking. As a result we have to revert to Vilenkin's Klein Gordon probability idea, in order to obtain a sensible probability definition. We did this in Section~\ref{wavepacks} and in Section~\ref{probinfl} evaluated the nucleation probability itself. 
We found that for small $\sigma_c$ we recover the plane wave results, with replacement (\ref{c2rep}). However, for large $\sigma_c$, higher order corrections could reverse this conclusion.

{\it Acknowledgments.} We would like to thank Steffen Gielen, Jonathan Halliwell and Alex Vilenkin
for discussions and advice in relation to this paper. This work was supported by the STFC Consolidated Grant ST/T000791/1 (JM),
ST/W507519/1 (EA) and The Simons Foundation.

\bibliographystyle{ieeetr}
\bibliography{bib}

\end{document}